\documentclass[aps,twocolumn,prl,superscriptaddress,longbibliography,reprint]{revtex4-2}
\usepackage{amsfonts}
\usepackage{mathrsfs}
\usepackage{amsmath}
\usepackage{color}
\usepackage{graphicx}
\usepackage{bm}
\usepackage{amssymb}
\usepackage{xspace}
\usepackage{epstopdf}
\usepackage{longtable}
\usepackage{multirow}
\usepackage{booktabs}
\usepackage{textcomp}
\usepackage[title]{appendix}
\usepackage{float}
\usepackage{dsfont}
\usepackage[colorlinks=true, letterpaper=true, pdfstartview=FitV, linkcolor=blue, citecolor=blue, urlcolor=blue]{hyperref}
\usepackage[]{changes}
\usepackage{overpic}
\usepackage{makecell}
\usepackage{adjustbox}

\providecommand{\U}[1]{\protect\rule{.1in}{.1in}}

\begin{document}
\title{Theory of In-Plane Orbital Magnetization with Layer Hybridization}


\author{Jin-Xin Hu}\thanks{jhuphy@ust.hk}
\affiliation{Department of Physics, The Hong Kong University of Science and Technology, Clear Water Bay, Hong Kong SAR, China}

\affiliation{Center for Theoretical Condensed Matter Physics, The Hong Kong University of Science and Technology, Clear Water Bay, Hong Kong SAR, China}

 \author{Zi-Ting Sun}\thanks{zsunaw@connect.ust.hk}

\affiliation{RIKEN Center for Emergent Matter Science (CEMS), Wako, Saitama 351-0198, Japan}

   \author{Yugui Yao}\thanks{ygyao@bit.edu.cn}
\affiliation{Centre for Quantum Physics, Key Laboratory of Advanced Optoelectronic Quantum Architecture and Measurement (MOE), School of Physics, Beijing Institute of Technology, Beijing, 100081, China}

\begin{abstract}
The modern theory of orbital magnetization successfully describes the response of Bloch electrons to magnetic fields in fully periodic crystals, but it does not directly address the distinct regime of an in-plane field in multilayer systems with layer hybridization. Coherent interlayer tunneling allows electrons to form circulating current loops, producing an in-plane orbital response that is absent in a strictly two-dimensional limit and qualitatively different from the conventional three-dimensional one. Here we develop a theory of in-plane orbital magnetization for this {\it transdimensional} regime, where the layer thickness is comparable to the vertical mean free path. Starting from the current-loop picture, we construct the in-plane orbital angular momentum operator and derive exact expressions for the orbital magnetic moment and the in-plane orbital magnetic susceptibility. As an application, we predict a gate-tunable in-plane orbital magnetoelectric effect in layered materials. Our framework establishes a general foundation for in-plane orbital responses and suggests new opportunities for orbitronics in layer-hybridized quantum materials.

\end{abstract}
\maketitle

\emph{Introduction.}---Orbital magnetization is a fundamental thermodynamic property of electrons in solids and has become an essential concept in modern quantum materials~\cite{getzlaff2007fundamentals,hirst1997microscopic,vanderbilt2018berry,bernevig2022progress,mei2024electrically,grytsiuk2020topological,he2022topological,zhong2016gyrotropic}. In crystalline systems, the modern theory of orbital magnetization shows that the orbital response of Bloch electrons cannot, in general, be reduced to a local current density within a unit cell; instead, it also contains a Berry-phase contribution~\cite{thonhauser2005orbital,ceresoli2006orbital,resta2010electrical,bianco2013orbital,topp2022orbital,hanke2016role}. This framework has provided a unified understanding of orbital magnetization in periodic systems and clarified its relation to band geometry and topology~\cite{xiao2006berry,chang1996berry,chang2008berry,xiao2020unified,shi2007quantum}.

\begin{figure}
		\centering
		\includegraphics[width=0.9\linewidth]{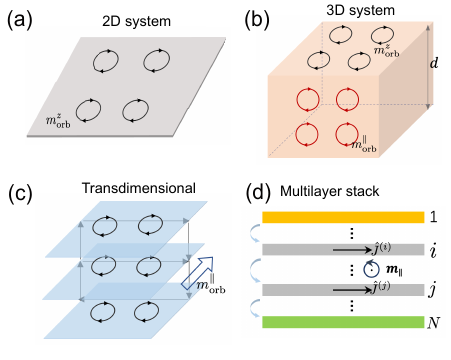}
		\caption{Dimensional crossover of orbital magnetic moments in layered systems. (a) In an ideal two-dimensional system, Bloch electrons carry only an out-of-plane orbital magnetic moment $m_{\mathrm{orb}}^z$. (b) In a three-dimensional crystal, both $m_{\mathrm{orb}}^z$ and an in-plane component $m_{\mathrm{orb}}^{\parallel}$ can occur. (c) In the transdimensional regime, $m_{\mathrm{orb}}^{\parallel}$ is generated by coherent interlayer orbital motion associated with circulating interlayer currents. (d) In a layer-hybridized multilayer stack, the current loop formed between the $i$-th and $j$-th layers contributes to the in-plane orbital magnetic moment.} 
		\label{fig:fig1}
\end{figure}

A distinct problem arises in layered materials subject to an in-plane magnetic field. Representative platforms include stacked van der Waals heterostructures such as few-layer graphene~\cite{han2023orbital,hu2026orbital,ghorai2025planar,li2026transdimensional} and moir\'{e} heterostructures (twisted graphene and transition metal dichalcogenides)~\cite{bistritzer2011moire,cao2018unconventional,cao2018correlated,lee2019theory,hu2024colossal,zhai2023time,wu2019topological,devakul2021magic,zheng2025layer}. In a strictly two-dimensional electron system, the orbital effect of an in-plane field is strongly suppressed because the Lorentz force does not generate the usual cyclotron motion within the plane. In multilayer systems, however, the combination of finite thickness and interlayer hybridization introduces an additional orbital channel: electrons can tunnel between layers and form closed circulating paths in real space. This circulating interlayer motion gives rise to an in-plane orbital magnetic response that is absent in the ideal two-dimensional limit.

The character of this orbital motion depends sensitively on spatial dimensionality. As illustrated in Fig.~\ref{fig:fig1} (a), an ideal two-dimensional system supports only an out-of-plane orbital moment $m^{z}_{\rm orb}$, whereas in a conventional three-dimensional system where the thickness $d$ is much larger than the vertical mean free path (coherence length) $l_z$, coherent interlayer orbital motion is suppressed by scattering, and the in-plane magnetization approximately satisfies $M_{\parallel} \approx M_{\text{3D}} \times (d / a_z)$, where $a_z$ is the interlayer spacing. In this case, the magnetization reduces to a thickness-averaged 3D response, as shown in Fig.~\ref{fig:fig1}(b). Between these two limits lies a transdimensional regime, where $d$ is comparable to $l_z$. In this regime, coherent interlayer orbital motion survives and produces a qualitatively distinct in-plane magnetization associated with circulating interlayer currents~\cite{li2026transdimensional}, as in Fig.~\ref{fig:fig1} (c). Despite growing interest in layered quantum systems~\cite{duong2017van,montblanch2023layered,turunen2022quantum,zhai2025twistronics,xiong2024antiscreening}, a general theory of in-plane orbital magnetic response in the transdimensional regime is still lacking.

In this work, we develop a systematic theory of in-plane orbital magnetization in layer-hybridized quantum systems. Within the current loop picture, we construct the in-plane orbital magnetic moment operator and confirm its validity by examining full quantum treatment based on Peierls substitution. Building on this framework, we further derive a complete formula for the in-plane orbital magnetic susceptibility and identify its geometrically distinct contributions~\cite{freimuth2017geometrical,piechon2016geometric,gao2015geometrical}. As a concrete physical application, we predict a gate-tunable in-plane orbital magnetoelectric (Edelstein) effect in time-reversal-symmetric layered materials. Our unified framework of in-plane orbital responses in van der Waals multilayers opens a route toward orbital functionalities enabled by coherent interlayer motion.

\emph{Theory of in-plane orbital magnetization.}---To describe the orbital responses of Bloch electrons in an extended periodic system, one may start from the antisymmetric form of orbital magnetic moment operator~\cite{song2019low,bhowal2021orbital} as
\begin{equation}
\hat{\bm{m}}=\frac{e}{4}(\hat{\bm{v}}\times \hat{\bm{r}}-\hat{\bm{r}}\times \hat{\bm{v}}),
\end{equation}
where $\bm{v}$ and $\bm{r}$ are velocity and position operators, respectively. Evaluating $\hat{\bm{m}}$ in the Bloch basis yields gauge-invariant expressions for orbital magnetic moments in ordinary periodic systems~\cite{xiao2010berry,shi2007quantum}. In a quasi-2D layered system, however, this form is not applicable because the out-of-plane momentum is not a good quantum number. As shown in Fig.~\ref{fig:fig1}(d), a general multilayer stack consists of electrons confined to individual layers. In such a system, electrons remain extended in the plane but can tunnel coherently between layers, so an in-plane orbital response is naturally associated with circulating interlayer current loops. This observation motivates a layer-resolved construction of the in-plane orbital magnetic moment operator as
\begin{equation}
\hat{\bm{m}}_\parallel = \frac{e}{4}\sum_{i\neq j}(\hat{\bm{J}}^{(i)}-\hat{\bm{J}}^{(j)})\times (\bm{Z}^{(i)}-\bm{Z}^{(j)}).
\end{equation}
Here $\hat{\bm{J}}^{(i)}$ denotes the current operator projected to the $i$-th layer and $\bm{Z}^{(i)}$ is the vertical position of the $i$-th layer. Each pair of layers ($i$ and $j$) contributes one circulating-current channel, and summing over $i\neq j$ collects the full in-plane orbital response. The projected current operator is $\hat{\bm{J}}^{(i)}=\frac{1}{2}\{\hat{\bm{v}},\hat{P}_i\}$, where $\hat{P}_i$ projects onto the Wannier orbitals belonging to layer $i$ ~\cite{zhou2024skin}. The operator $\hat{P}_i$ and position $\bm{Z}^{(i)}$ satisfy
\begin{equation}
\sum_{i}\hat{P}_i=I, \quad \sum_{i}Z^{(i)}=0.
\end{equation}
Using these relations, one directly finds $\sum_{i\neq j}(\hat{P}_i-\hat{P}_j)\times (Z^{(i)}-Z^{(j)})=4\sum_i \hat{P}_i Z^{(i)}$. It is then convenient to define the vertical position operator $\hat{\mathcal{Z}}=\sum_i \hat{P}_i Z^{(i)}$, in terms of which the in-plane orbital moment operator becomes
\begin{equation}
\label{eq:eq_oam}
\hat{m}^a=\frac{e}{2}\epsilon_{abz}(\hat{v}^b \hat{\mathcal{Z}}+\hat{\mathcal{Z}}\hat{v}^b).
\end{equation}
Here $a,b=x,y$. In the Bloch-band representation, the corresponding orbital magnetic moment for band $n$ is
\begin{equation}
\label{eq_eq_intrao}
m^a_n=e\epsilon_{abz} \sum_{m}\mathrm{Re}(v_{nm}^b \mathcal{Z}_{mn})
\end{equation}
The interband matrix elements are similarly given by $m^a_{nn'}=\langle n |\hat{m}_a|n'\rangle$, where $|n \rangle \equiv\left|u_{n \boldsymbol{k}}\right\rangle$ is the cell-periodic part of the Bloch eigenstate. The static in-plane orbital magnetization then follows as
\begin{equation}
M_{\mathrm{orb}}^{a}=\sum_{n}\int_{\bm{k}} \frac{d^2\bm{k}}{(2\pi)^2}m^a_n f_{n\bm{k}}.
    \label{eq:pom}
\end{equation}
where $f_{n\bm{k}}\equiv f(\varepsilon_{n\bm{k}})$ is the Fermi-Dirac distribution and $\varepsilon_{n\bm{k}}$ is the eigen-energy of the Bloch Hamiltonian $H_0(\boldsymbol k)$. A static in-plane orbital magnetization is symmetry-allowed only in magnetic point groups that admit a time-reversal-odd axial vector within the basal plane. For instance, the $x$-component requires the breaking of rotation $C_{nz}$ and mirror $\mathcal{M}_y$ symmetries.

\emph{Quantum verification from Peierls substitution}: We now show that the current-loop construction above can be reproduced by a full quantum treatment. We consider the perturbed Hamiltonian $H_0(\bm{k},\bm{B})=H_0(\bm{k})-\hat{\bm{m}}_\parallel \cdot \bm{B}_\parallel$ with $\bm{B}=(B_x,B_y)$ and decompose the Bloch Hamiltonian into layer blocks as
\begin{equation}
    H_0(\boldsymbol k)
    =
    \sum_{ij} H_{ij}(\boldsymbol k),
    \quad
    H_{ij}(\boldsymbol k)
    =
    \hat{P}_i H_0(\boldsymbol k) \hat{P}_j.
\end{equation}
Applying an in-plane magnetic field $\boldsymbol B_\parallel$ and choosing the gauge $\boldsymbol A(Z^{(i)})=Z^{(i)}(B_y,-B_x,0)$, the minimal coupling for the $ij$ layer block reads \cite{lee2019theory}
\begin{equation}
    H_{ij}(\boldsymbol k)
    \rightarrow
    H_{ij}
    \left[
    \boldsymbol k+
    \frac{e}{\hbar}\frac{Z^{(i)}+Z^{(j)}}{2}
(B_y,-B_x)    \right].
    \label{eq:block_minimal_coupling}
\end{equation}
Expanding Eq.~\eqref{eq:block_minimal_coupling} to linear order in $\boldsymbol B_\parallel$ and summing over all layer blocks, we obtain the corresponding orbital moment operator follows from differentiating the field-dependent Hamiltonian: $\hat{m}_a=-\partial H(\bm{k},\bm{B})/\partial B_a|_{\bm{B}=0}$, which gives
\begin{equation}
\hat{m}_a=e\epsilon_{abz}\sum_{i,j}\hat{P}_i \frac{Z^{(i)}+Z^{(j)}}{2}\hat{v}^b_{ij}\hat{P}_j , 
\end{equation}
in which $\hat{v}^b_{ij}=\partial H_{ij}(\bm{k})/\partial k_b$. Using $\hat{v}^b_{ij}= \hat{P}_i \hat{v}^b \hat{P}_j$, we recover Eq.~\eqref{eq:eq_oam} exactly. This agreement shows that the layer-resolved current-loop construction is not merely heuristic but is fully consistent with the quantum approach.

\emph{Relation to the modern theory of orbital magnetization.}---Having established the operator construction, we next clarify how the present theory connects to the modern theory of orbital magnetization~\cite{thonhauser2005orbital,xiao2010berry}. According to that theory, the orbital magnetization consists 
of two gauge-invariant contributions: a local orbital 
magnetic moment (self-rotation of Bloch electrons) and a Berry curvature term (center-of-mass motion). To see this 
decomposition, we examine an effective field-dependent free energy
\begin{equation}
\label{eq:eq_freeb}
F(\bm{B}) = -\frac{1}{\beta} \sum_{n\bm{k}} 
\left(1 + \frac{e}{\hbar} \bm{B}\cdot\bm{\Omega}_{n \bm{k}}\right)
\phi\bigl[\varepsilon_{n\mathbf{k}}(\bm{B})\bigr],
\end{equation}
where $\phi(\varepsilon) = \ln[1+e^{-\beta(\varepsilon-\mu)}]$, 
and the field-dependent band energy is 
$\varepsilon_{n\bm{k}}(\bm{B}) = \varepsilon_{n\mathbf{k}}^{(0)} 
- \bm{m}_{n\mathbf{k}}\cdot\bm{B}$. Here $\bm{m}_{n\bm{k}}$ and $\bm{\Omega}_{n\bm{k}}$ are the orbital magnetic moment and Berry curvature, respectively. In layered systems with the vertical position operator $\mathcal{Z}$, the relevant quantities become $\tilde{\bm{m}}_{n\bm{k}} = e \sum_{m\neq n} 
\bm{v}_{nm} \times \bm{\mathcal{Z}}_{mn}$ and 
$\tilde{\bm{\Omega}}_{n\bm{k}} = \nabla_{\bm{k}} 
\times \bm{\mathcal{Z}}_{n}$~\cite{ghorai2025planar}. Substituting these into the free energy and expanding to first order in 
$\bm{B}$, we obtain two distinct terms of the free energy $F=F_1+F_2$, which reads
\begin{align}
F_1 &= \frac{1}{\beta} \sum_{n\bm{k}} \ln\Bigl[1 + 
e^{-\beta(\varepsilon_{n\mathbf{k}}^{(0)} - 
\tilde{\bm{m}}_{n\bm{k}}\cdot\bm{B} - \mu)}\Bigr],\\
F_2 &= -e \bm{B} \cdot \sum_{n\bm{k}} 
\bm{v}_{n\mathbf{k}} \times \bm{\mathcal{Z}}_{n} 
 f_{n\bm{k}}.
\end{align}
The total orbital magnetization is $M_{\text{orb}}^a = -\partial_{B_a}(F_1+F_2)|_{\bm{B}=0}$, consistent with Eq.~\eqref{eq_eq_intrao}. We thus identify two gauge-invariant contributions to the in-plane orbital response in layered materials: an interband contribution (the sum over $m\neq n$) associated with virtual coherence between bands and an intraband layerflow contribution (the $m=n$ term) inherited from the Berry-phase structure. In the present layered setting, the Berry-curvature part of the orbital magnetization from modern theory naturally maps onto $\bm{v}_{n} \times \boldsymbol{\mathcal{Z}}_{n}$.

\begin{figure}
		\centering
		\includegraphics[width=1.0\linewidth]{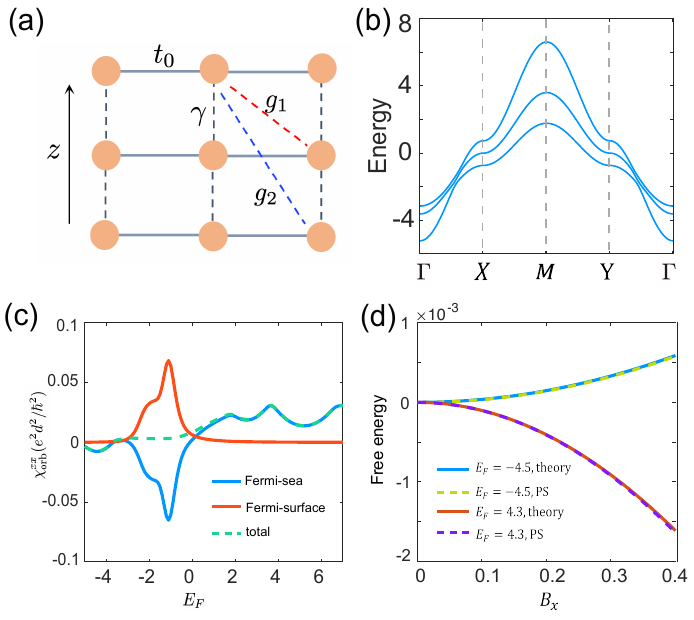}
		\caption{Model validation of the in-plane orbital magnetic susceptibility. (a) Three-layer square-lattice model with intralayer hopping $t_0$ and interlayer hoppings $\gamma$, $g_1$, and $g_2$. (b) Band structure for the model used in the numerical calculations. (c) In-plane orbital susceptibility $\chi_{\mathrm{orb}}^{xx}$ as a function of Fermi energy $E_F$, decomposed into Fermi-surface, Fermi-sea, and total contributions. (d) Field-dependent free-energy shift under an in-plane magnetic field $B_x$. Dashed curves are obtained from the full Peierls-substitution calculation, while solid curves show $-\frac{1}{2}\chi_{\mathrm{orb}}^{xx}B_x^2$ from Eq.~\eqref{eq:planar_orb_sus_final}. Parameters are $U_d=0.2$ and $k_B T=0.1$.} 
		\label{fig:fig2}
\end{figure}

\emph{In-plane orbital magnetic susceptibility.}---We now turn to the second-order magnetic response, which provides a compact description of the in-plane orbital physics beyond the static moment itself. The orbital magnetic susceptibility is defined through the second-order correction to the free energy, which reads
\begin{equation}
F=F_0-\bm{M}_{\mathrm{orb}}\cdot\bm{B}-\frac{1}{2}\chi_{\mathrm{orb}}^{ab}B_a B_b.
\end{equation}
Equivalently, $\chi_{\mathrm{orb}}^{ab}=\partial M^a_{\mathrm{orb}}/\partial B_b$. If the system has \(C_{3z}\), or higher, rotational symmetry, this tensor is
isotropic in the plane: $\chi^{xx}=\chi^{yy}$ and $\chi
^{xy}=\chi^{yx}=0$. We can derive the generic expression as
\begin{equation}
\chi_{\rm orb}^{ab}
=-\sum_n
\int_{\bm{k}}\frac{d^2\bm{k}}{(2\pi)^2}\left\{
f_n'
m_n^am_n^b+f_n\mathcal C^{ab}_n\right\}.
\label{eq:planar_orb_sus_final}
\end{equation}
The two terms in Eq.~\eqref{eq:planar_orb_sus_final} have transparent
physical meanings. The former terms is a Fermi-surface contribution.
Since \(f_n'\le 0\), it gives a Van Vleck-type paramagnetic contribution and describes the redistribution of pre-existing planar orbital
moments near the Fermi surface under the first-order energy correction $\delta\varepsilon_{n\bm{k}}(\bm{B}_\parallel)=-\bm{m}_n\cdot \bm{B}$.

The latter Fermi-sea contribution comes from the $\bm{B}^2_\parallel$ correction to the band energy,
    $\delta\varepsilon_{n\boldsymbol k}(\bm{B}^2_\parallel)
    =\frac12
    B_aB_b
    \mathcal C_n^{ab}$, which reads
\begin{equation}
    \mathcal C_n^{ab}=2\sum_{l\neq n}\frac{\mathrm{Re}[m_a^{nl}m_b^{ln}]}{\varepsilon_n-\varepsilon_l}+(\hat{h}_{ab})_{nn}.
    \label{eq:Cn_ab_sus}
\end{equation}
The first term in $\mathcal C_n^{ab}$ encodes the interband geometric property, which originates from the virtual magnetic mixing between different Bloch bands through the off-diagonal matrix elements of $\bm{m}_\parallel^{nl}$, and remains finite in a band insulator. The second term comes from the Langevin-type diamagnetic correction of $H_0(\boldsymbol k)$ due to $\boldsymbol B_\parallel$ as $\delta H^{\rm orb}_d= -\hat{h}_{\mu \nu} B_\mu B_\nu$, where $h_{ab}$ is given by
\begin{equation}
h_{\mu \nu}=\frac{1}{2}e\epsilon_{\mu \nu z}\{ \hat{m}_{\mu \nu},\hat{\mathcal{Z}}\}
\end{equation}
where $\hat{m}_{ab}=e\epsilon_{ac z}\{ \partial^2H_0/\partial k_a\partial k_b, \hat{\mathcal{Z}}\}/(2\hbar^2)$. Importantly, this term can be rewritten as $h_{ab}=-e\mathcal D_b\mathcal Q^{\rm orb}_{za}$, which identifies a diamagnetic vertex associated with a planar orbital magnetic quadrupole. Here $\mathcal D_b
    =
    \left(
    \hat{z}
    \times
    \boldsymbol\nabla_{\hbar\boldsymbol k}
    \right)_b$ and
\begin{equation}
    \mathcal Q^{\rm orb}_{za}
    \equiv
    \sum_i
   Z^{(i)}\hat{m}_a^i.
    \label{eq:Qza_layer}
\end{equation}
Here $\hat{m}_a^i=\left\{ P_i, \hat{m}_a \right\}/2$ is the layer-resolved in-plane orbital moment. Therefore, $\mathcal Q^{\rm orb}_{za}$ measures the $z$-distribution of the in-plane orbital moment.

\emph{Model Hamiltonian and validation.}---To illustrate the theory and validate the susceptibility formula, we consider an $N$-layer square-lattice model. Under the basis $c_{\bm{k}}=(c_{1\bm{k}},c_{2\bm{k}},...,c_{N\bm{k}})^T$, the non-interacting Hamiltonian is $H_0=\sum_{\bm{k}}c_{\bm{k}}^\dagger H_0(\bm{k})c_{\bm{k}}$, where $H_0(\bm{k})$ reads
\begin{equation}
H_0(\bm{k}) = \begin{pmatrix}
 h_\ell^{(0)} &  h^{(1)} &  h^{(2)} & 0 & \cdots \\
 h^{(1)^\dagger} &  h_\ell^{(0)} &  h^{(1)} &  h^{(2)} & \ddots \\
h^{(2)^\dagger} & h^{(1)^\dagger} & h_\ell^{(0)} & h^{(1)} & \ddots \\
0 & h^{(2)^\dagger} & h^{(1)^\dagger} & h_\ell^{(0)} & \ddots \\
\vdots & \ddots & \ddots & \ddots & \ddots
\end{pmatrix},
\label{eq:eq_Hamil}
\end{equation}
where $h_\ell^{(0)}=-2t_0(\cos k_x+\cos k_y)+u_{\ell} $, $h^{(1)}=\gamma-2g_1(\cos k_x+\cos k_y)$, and $h^{(2)}=2g_2(\cos k_x+\cos k_y)$. We take parameters $(t_0,\gamma,g_1,g_2)=(1,0.5,0.3,0.1)$ throughout this work. Finally, $u_{\ell}$ are on-site potentials induced by an external displacement field. For simplicity we consider a potential difference $U_d$ between the top and bottom layers with $u_{\ell} = U_d \left(\frac{1}{2}+\frac{1-\ell}{N-1} \right)$. Figure~\ref{fig:fig2}(a) shows the resulting three-layer geometry, and Fig.~\ref{fig:fig2}(b) displays the corresponding band structure along the high symmetry line.

We first evaluate $\chi_{\mathrm{orb}}^{xx}$ as a function of the Fermi energy, as shown in Fig.~\ref{fig:fig2}(c). The paramagnetic part arises entirely from the Fermi-surface contribution, whereas the diamagnetic part contains both the virtual interband term and the orbital magnetic quadrupole contribution discussed above. We further validate the susceptibility formula by computing the free energy as a function of the in-plane magnetic field $B_x$ using the quantum Peierls substitution (Eq.~\eqref{eq:block_minimal_coupling}). The dashed curves in Fig.~\ref{fig:fig2}(d) are obtained from the numerical free energy, while the solid curves represent $-\frac{1}{2}\chi_{\mathrm{orb}}^{xx}B_x^2$ (Eq.~\eqref{eq:planar_orb_sus_final}). Their agreement confirms our analytical expression for the in-plane orbital magnetic susceptibility.

\begin{figure}
		\centering
		\includegraphics[width=1.0\linewidth]{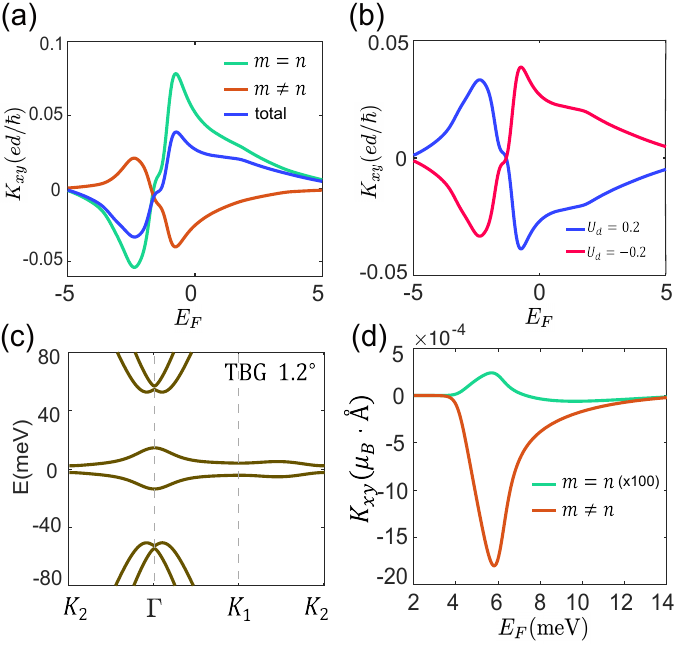}
		\caption{In-plane orbital magnetoelectric response. (a) Kinetic magnetoelectric coefficient $K_{xy}$ as a function of Fermi energy $E_F$, resolved into intraband ($m=n$), interband ($m\neq n$), and total contributions. (b) Gate-controlled switching of $K_{xy}$ under opposite displacement fields $U_d=\pm 0.2$. (c) Representative moire miniband structure of twisted bilayer graphene at twist angle $1.2^\circ$. (d) Calculated $K_{xy}$ in the twisted-bilayer-graphene model, showing the relative intraband and interband contributions.}
		\label{fig:fig3}
\end{figure}

\emph{In-plane magnetoelectric (Edelstein) effect.}---The in-plane orbital motion of Bloch electrons directly allows current-induced magnetization. Here we examine the in-plane magnetoelectric (Edelstein) effect and show how it can be controlled by a gate electric field in layered materials. In magnetoelectric effects, the induced magnetization $\bm{M}$ and the applied electric field $\bm{E}$ are related by the magnetoelectric pseudotensor $\alpha$ such that $M_i = \sum_j \alpha_{ij} E_j$, where $i,j = x,y$ and $\alpha_{ij}$ are the elements of $\alpha$. The tensor $\alpha_{ij}$ can be further written as
\begin{equation}
\alpha_{ij}= e\int_{n\bm{k}}\tau m_n^i\partial_jf_{n\bm{k}}+\sum_{l\neq n}\frac{2\mathrm{Re}(m_{nl}^ir_{ln}^j)}{\varepsilon_{nl}}f_{n\bm{k}},
\end{equation}
where $\alpha_{ij} = \alpha_{ij}^{\text{ex}} + \alpha_{ij}^{\text{in}}$ contains the extrinsic and intrinsic contributions. $\tau$ denotes the relaxation time. In Table~\ref{tab:KPOM_symmetry_compact} we present the crystalline symmetry constraints for in-plane component $\alpha_{ij}$. Importantly, $\alpha_{ij}^{\text{ex}}$ survives $\mathcal{T}$ symmetry while $\alpha_{ij}^{\text{in}}$ survives $\mathcal{PT}$ symmetry. Unlike the out-of-plane magnetoelectric response, the in-plane component $\alpha_{ij}$ is nonzero for systems with $C_{3z}$ or higher rotational symmetries.


\begin{table}[t]
\centering
\scriptsize
\setlength{\tabcolsep}{2.2pt}
\renewcommand{\arraystretch}{1.15}
\caption{Symmetry constraints on the in-plane orbital magnetoelectric response under crystal symmetry operation $\mathcal{R}$. A check mark means that the symmetry does not forbid the response, while a cross means that the response is forbidden if the symmetry is present.}
\label{tab:KPOM_symmetry_compact}
\begin{ruledtabular}
\begin{tabular}{c|ccccc}
 & \({\cal P}\) & \(C_{nz}\) & \(\mathcal{M}_x/C_{2x}\) & \(\mathcal{M}_y/C_{2y}\) & \(\mathcal{M}_z\)  \\
\hline
\(\alpha_{xx}\) & \(\times\) & \(\checkmark\) & \(\times\) & \(\times\) & \(\times\)  \\
\(\alpha_{yy}\) & \(\times\) & \(\checkmark\) & \(\times\) & \(\times\) & \(\times\)  \\
\(\alpha_{xy}\) & \(\times\) & \(\checkmark\) & \(\checkmark\) & \(\checkmark\) & \(\times\) 
\end{tabular}
\end{ruledtabular}
\end{table}

The numerical results for the in-plane orbital magnetoelectric effect are shown in Fig.~\ref{fig:fig3}. Here we focus on $\mathcal{T}$-symmetric systems and write $\alpha_{ij}^{\text{ex}} = e\tau K_{ij}$, where $K_{ij}$ is the kinetic magnetoelectric susceptibility~\cite{zhong2016gyrotropic,wang2024spontaneous}. For the model Hamiltonian introduced above, $K_{xy}$ is evaluated and plotted in Fig.~\ref{fig:fig3}(a) as a function of the Fermi energy $E_F$. We decompose $K_{xy}$ into interband coherence and intraband layerflow contributions and find that both terms contribute to the total response. Notably, $K_{xy}$ changes sign when the gate electric field $U_d$ is switched, indicating that the generation of orbital magnetization is locked to the polarization associated with the position operator $\mathcal{Z}$. We also consider another system: a twisted bilayer graphene (TBG) encapsulated by hBN. The moir\'e bands at twist angle $\theta=1.2^\circ$ are shown in Fig.~\ref{fig:fig3}(c), and the corresponding $K_{xy}$ is calculated in Fig.~\ref{fig:fig3}(d). In this case, the interband coherence term dominates over the intraband term, implying that the weak band velocity $\bm{v}_n$ in a flat-band system gives only a small contribution to the in-plane magnetoelectric response.



\emph{Conclusion and discussion.}---In summary, we have formulated a compact theory of in-plane orbital magnetization in layered quantum systems. Crucially, we derive exact expressions for both the first-order orbital magnetic moment operator and the second-order orbital magnetic susceptibility in the transdimensional regime when the interlayer coherent motion plays the dominant role.

Our general framework has several immediate physical implications. First, the Berry connection for a given band $n$ acquires a gauge-invariant correction 
$\delta A_n^i = \Gamma_{ij} B_j$, where
\begin{equation}
\Gamma_{ij} = \sum_{l \neq n} 2 \operatorname{Re} 
\left( \frac{r_{ln}^i \, m_{nl}^j}{\varepsilon_{ln}} \right),
\end{equation}
which describes how a Bloch electron acquires a field-induced positional shift in response to an in-plane magnetic field~\cite{gao2014field}. This position shift accounts for intrinsic responses related to band geometric effects of Bloch electrons~\cite{gao2014field,liu2021intrinsic,liu2022berry,wang2024orbital,huang2023intrinsic,fang2024quantum,kaplan2024unification,xiang2023third,lai2021third}.

Second, the orbital magnetic susceptibility can be directly used to derive nonlinear responses to an in-plane magnetic field. Consider an external perturbation that couples to a local operator via $\hat{\bm{\mathcal{O}}} \cdot \bm{f}$. The second-order correction to the observable $\delta \langle \hat{\mathcal{O}} \rangle_c = 
\kappa_{ab}^c B_a B_b$ is determined by the susceptibility tensor $\kappa_{ab}^c$. From the free energy expansion and thermodynamical relation, we can obtain 
\begin{equation}
\kappa_{ab}^c = \frac{1}{2}\left. \frac{\partial \chi_{\mathrm{orb}}^{ab}}
{\partial f_c} \right|_{\bm{f}=0}.
\end{equation}
This relation connects the orbital magnetic susceptibility to higher-order response functions, providing a route to probe geometric effects via nonlinear magneto-transport measurements.



%

\clearpage
		\onecolumngrid
\begin{center}
		\textbf{\large Supplementary Material for\\ ``Theory of In-Plane Orbital Magnetization with Layer Hybridization''}\\[.2cm]

\end{center}
	
	\maketitle

\setcounter{equation}{0}
\setcounter{section}{0}
\setcounter{figure}{0}
\setcounter{table}{0}
\setcounter{page}{1}
\renewcommand{\theequation}{S\arabic{equation}}
\renewcommand{\thesection}{ \Roman{section}}

\renewcommand{\thefigure}{S\arabic{figure}}
\renewcommand{\thetable}{\arabic{table}}
\renewcommand{\tablename}{Supplementary Table}

\renewcommand{\bibnumfmt}[1]{[S#1]}
\renewcommand{\citenumfont}[1]{#1}
\makeatletter

\maketitle

\setcounter{equation}{0}
\setcounter{section}{0}
\setcounter{figure}{0}
\setcounter{table}{0}
\setcounter{page}{1}
\renewcommand{\theequation}{S\arabic{equation}}
\renewcommand{\thesection}{ \Roman{section}}

\renewcommand{\thefigure}{S\arabic{figure}}
\renewcommand{\thetable}{\arabic{table}}
\renewcommand{\tablename}{Supplementary Table}

\renewcommand{\bibnumfmt}[1]{[S#1]}
\renewcommand{\thesection}{S\arabic{section}}
\renewcommand{\theequation}{S\arabic{equation}}
\renewcommand{\thetable}{S\arabic{table}}
\renewcommand{\thefigure}{S\arabic{figure}}
\setcounter{equation}{0}
\setcounter{page}{1}

\maketitle

\makeatletter 

\section{\bf{Orbital magnetic moment in two-band model}}
In the main text, we have shown the proper construction of the in-plane orbital magnetic moment operator. It is instructive to give an analytical formula for an effective two-band model with the basis $|t\rangle$ and $|b\rangle$. Here $t,b$ denote the top layer and bottom layer, respectively. Recall the orbital magnetic moment for a given band $n$ as
\begin{equation}
\label{eq_eq_intrao}
m_a^n=e\epsilon_{abz} \sum_{m}\mathrm{Re}(v_{nm}^b \mathcal{Z}_{mn})
\end{equation}
We consider a simple model with effective bilayer, such as rhombohedral-stacked multilayer graphene. A general two-band model reads
\begin{equation}
H=h_0(\bm{k})\sigma_0+h_x(\bm{k})\sigma_x+h_y(\bm{k})\sigma_y+h_z(\bm{k})\sigma_z,
\end{equation}
where $\sigma$ acts on the layer-pseudospin space. The eigenstates read
\begin{equation}
|c\rangle=\left(\begin{array}{c}\cos\frac{\theta}{2} \\
e^{i\varphi}\sin\frac{\theta}{2}
\end{array}\right), |v\rangle=\left(\begin{array}{c}
\sin\frac{\theta}{2} \\
-e^{i\varphi}\cos\frac{\theta}{2}
\end{array}\right),
\end{equation}
where $\cos\theta=h_z/|\bm{h}|$ and $\tan\varphi=h_y/h_x$. $c,v$ denotes the conduction and valence band. The vertical position operator is $\mathcal{Z}=d\sigma_z/2$, where $d$ is the interlayer distance for the target bilayer system.

After some calculations, we can derive the in-plane orbital magnetic moment as
\begin{equation}
m_a=\pm \epsilon_{abz}\frac{ed}{2\hbar}(\partial_b h_z+\cos\theta \partial_bh_0).
\end{equation}
Here $\pm$ denotes the conduction and valence band. Importantly, the first term is the interband coherence ($m\neq n$)  term, while the second term is the intraband ($m=n$) term.

\section{\bf{Details of calculation for TBG}}
We provide the calculations of the band structure as well as the in-plane magnetic responses for twisted bilayer graphene (TBG). TBG is formed by two layers with a twist angle $-\theta/2$ and $\theta/2$ relative to $x$ axis respectively. In the continuum limit by neglecting the intervalley mixing, the Hamiltonian reads
\begin{equation}
H=H_1+H_2+H_{int}.
\end{equation}
Here, $H_{1/2}$ denotes the Hamiltonian of the top/bottom layer, and $H_{int}$ denotes the interlayer coupling. $H_{1/2}$ reads:
\begin{align}
H_{1} & =\sum_{\xi,\bm{k}}a^\dagger_{1,\xi,\bm{k}}\hbar v_F \hat{R}_{\theta/2}(\bm{k}-\bm{K_1})\cdot(\xi \sigma_x,\sigma_y)a_{1,\xi,\bm{k}}, \\ 
H_{2} & =\sum_{\xi,\bm{k}}a^\dagger_{2,\xi,\bm{k}}\hbar v_F \hat{R}_{-\theta/2}(\bm{k}-\bm{K_2})\cdot(\xi \sigma_x,\sigma_y)a_{2,\xi,\bm{k}},
\end{align}
where $\hat R_{\theta} = \begin{pmatrix} 
\cos \theta & -\sin \theta \\
\sin \theta & \cos \theta 
\end{pmatrix}$ is the rotation operator. $\sigma$ acts on the spinor of sublattices A and B in graphene. $\xi=\pm$ denotes the $K$ and $K'$ valleys. The interlayer coupling can be written as 
\begin{equation}
H_{int}=\sum_{\xi}\int_{\bm{r}}\psi^\dagger_{1,\xi}(\bm{r})T(\bm{r})\psi_{2,\xi}(\bm{r})+h.c.~,
\end{equation}
where 
\begin{equation}
T(\bm{r})=\left(
                       \begin{array}{cc}
                         w_0 & w_1 \\
                         w_1 & w_0 \\
                       \end{array}
                     \right)
                 + \left(
                       \begin{array}{cc}
                         w_0 & w_1e^{-i\frac{2\pi}{3}} \\
                         w_1e^{i\frac{2\pi}{3}} & w_0 \\
                       \end{array}
                     \right)e^{-i\bm{G}_1^M\cdot \bm{r}}
                  +   \left(
                       \begin{array}{cc}
                         w_0 & w_1e^{i\frac{2\pi}{3}} \\
                         w_1e^{-i\frac{2\pi}{3}} & w_0 \\
                       \end{array}
                     \right)e^{-i(\bm{G}_1^M+\bm{G}_2^M)\cdot \bm{r}}   ~.
\end{equation}
With the relation $\psi_{i,\xi}(\bm{r})=\sum_{\bm{k}}a_{i,\xi,\bm{k}}e^{i\bm{k}\cdot\bm{r}}$, the interlayer coupling can be expressed as 
\begin{equation}
H_{int}=\sum_{\xi}a^\dagger_{1,\xi,\bm{k}}[\left(
                       \begin{array}{cc}
                         w_0 & w_1 \\
                         w_1 & w_0 \\
                       \end{array}
                     \right)\delta_{\bm{k},\bm{k}'}+\left(
                       \begin{array}{cc}
                         w_0 & w_1e^{-i\frac{2\pi}{3}} \\
                         w_1e^{i\frac{2\pi}{3}} & w_0 \\
                       \end{array}
                     \right)\delta_{\bm{k+\bm{G}_1^M},\bm{k}'}+ \left(
                       \begin{array}{cc}
                         w_0 & w_1e^{i\frac{2\pi}{3}} \\
                         w_1e^{-i\frac{2\pi}{3}} & w_0 \\
                       \end{array}
                     \right)\delta_{\bm{k}+\bm{G}_1^M+\bm{G}_2^M,\bm{k}'}]a_{2,\xi,\bm{k}'}+h.c.~.
\end{equation}

\begin{figure}[t]
\centering 
\includegraphics[width=1\linewidth]{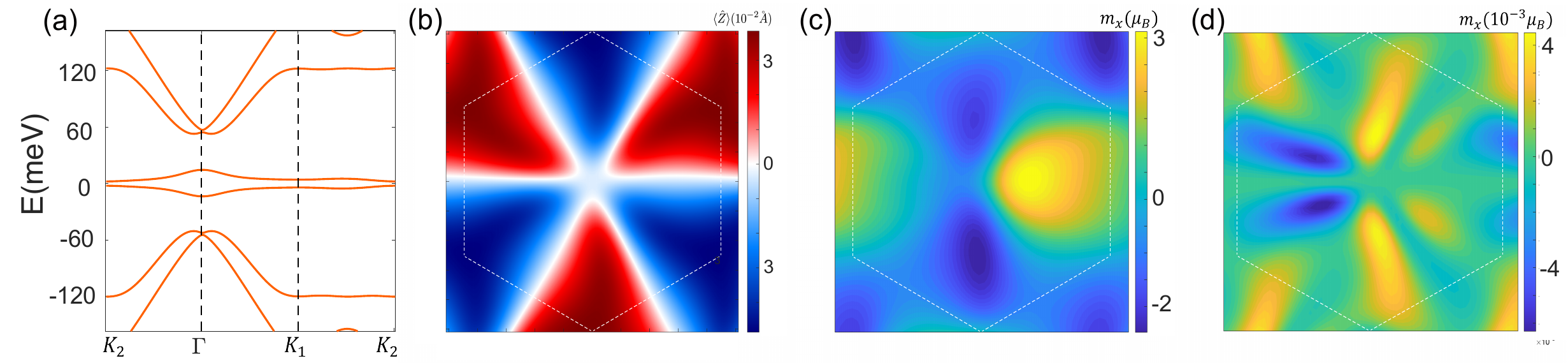}
\caption{(a) The moir\'{e} band structure of TBG at $\theta=1.2^\circ$. (b) The expectation of position operator $\langle \mathcal{Z}\rangle$ for the lowest conduction band. (c) The in-plane orbital magnetic moment. (d) The intraband layerflow term of $m_x$ (only $m=n$ term).}
\label{fig:figS1} 
\end{figure}

We adopt the parameters with $\hbar v_F=2365 a_0$ meV$\cdot\AA$, $w_0=80$ meV and $w_1=110$ meV. We add a sublattice potential $\Delta\sigma_z$ with $\Delta=15$ meV to the bottom layer when we consider the coupling with hBN substrate. In Fig.~\ref{fig:figS1}(a) and (b) we plot the band structure of TBG at $\theta=1.2^\circ$ and the corresponding vertical position $\mathcal{Z}_n=\langle \mathcal{Z}\rangle$ of the lowest conduction band. It is clear that the intraband term $\mathcal{Z}_n$ is small, implying that the interband coherence term $\mathcal{Z}_{nm}$ dominate. This can be seen in the numerical calculation of the orbital magnetic moment $m_{x}$ as shown in Fig.~\ref{fig:figS1}(c) and (d), where we plot the two contributions to see it more clearly.

\section{\bf{In-plane  magnetoresistance}}
Magnetoresistance, the change in electrical resistance of a material in response to an applied magnetic field, is one of the most fundamental transport phenomena in condensed matter physics. We now derive the in-plane magnetoresistance arising from the orbital geometric responses developed in our work. The Drude formula of the electric conductivity reads
\begin{equation}
\sigma_{aa}=-e^2\tau\sum_n\int \frac{d^2\bm{k}}{(2\pi)^2}v_{n\bm{k}}^2 f'_{n\bm{k}}=-e^2\tau\sum_n\int \frac{d^2\bm{k}}{(2\pi)^2}v_{n\bm{k}} \partial_{\bm{k}}f_{n\bm{k}}.
\end{equation}
In the presence of in-plane magnetic field, the band energy dispersion is modified to $\tilde{\varepsilon}_{n}=\varepsilon_n-\bm{m}_{n}\cdot \bm{B}$, and the group velocity is modified to $\tilde{\bm{v}}_{n}=\bm{v}_n-\partial_{\bm{k}}\bm{m}_{n}\cdot\bm{B}$. With these relations, we can write down the in-plane magnetoresistance as 
\begin{equation}
\sigma_{xx}(B)=\sigma_{xx}+\Delta \sigma B_a^2
\end{equation}
where the $\Delta \sigma$ can be obtained directly as
\begin{equation}
\Delta \sigma=e^2\tau \sum_n\int \frac{d^2\bm{k}}{(2\pi)^2}\partial^2_a m_{n\bm{k}}^a\cdot m_{n\bm{k}}^af'_{n}.
\end{equation}
This general formula can be employed to calculate the magnetoresistance in any layered materials. 

\section{\bf{Vertical position operator}}
For a general multilayer system, the vertical position operator is generally given by
\begin{equation}
\hat{\mathcal{Z}}=\sum_{\ell,\alpha} \ell  |\phi_{\ell,\alpha} \rangle \langle \phi_{\ell,\alpha} |
\end{equation}
where $ |\phi_{\ell,\alpha} \rangle$ denotes the $\alpha$ orbital on the layer at height $\ell$ measured from the center of the stack. To be specific, the matrix form of the operator for several examples can be written as
\begin{equation}
\label{eq:different_mag}
\hat{\mathcal{Z}}/d=\left\{
\begin{aligned}
 &\frac{1}{2}\mathrm{diag}(1,-1)  \quad (\mathrm{bilayer}) \\   
  &\frac{1}{2}\mathrm{diag}(1,0,-1) \quad (\mathrm{trilayer})  \\  
   &\frac{1}{2}\mathrm{diag}(1,\frac{1}{3},-\frac{1}{3},-1) \quad (\mathrm{tetralayer}) \\
   &\frac{1}{2}\mathrm{diag}(1,\frac{1}{2},0,-\frac{1}{2},-1) \quad (\mathrm{pentalayer})
\end{aligned}
\right.
\end{equation}
If additional orbital degrees of freedom (such as the sublattice index $\alpha=A,B$ in graphene) are considered, the expression for layer polarization operator can be readily modified. Here $d$ is the total separation between the top andbottom layers.

\end{document}